%% file: ns2_module_channel_access.tex
\documentclass[twocolumn]{article}
\usepackage{graphicx}
\usepackage{color}
\usepackage{url}
\usepackage{booktabs}
\usepackage{ulem}
\usepackage{authblk}
\usepackage{geometry}

\title{Physical Channel Access (PCA): Time and Frequency Access Methods Emulation in NS-2}

\author[1,2]{Nicolas Kuhn}
\author[1]{Olivier Mehani}
\author[2]{Huyen-Chi Bui}
\author[2]{J\'{e}r\^{o}me Lacan}
\author[2]{Jos\'{e} Radzik}
\author[2]{Emmanuel Lochin}
\affil[1]{NICTA, Sydney, Australia}
\affil[2]{University of Toulouse, ISAE, TeSA, France}

\begin{document}


\maketitle

\begin{abstract}
We present an NS-2 module, Physical Channel Access (PCA), to simulate different access methods on a link shared with Multi-Frequency Time Division Multiple Access (MF-TDMA). This technique is widely used in various network technologies, such as satellite communication. In this context, different access methods at the gateway induce different queuing delays and available capacities, which strongly impact transport layer performance. Depending on QoS requirements, design of new congestion and flow control mechanisms and/or access methods requires evaluation through simulations. 

PCA module emulates the delays that packets will experience using the shared link, based on descriptive parameters of lower layers characteristics. Though PCA has been developed with DVB-RCS2 considerations in mind (for which we present a use case), other MF-TDMA-based applications can easily be simulated by adapting input parameters. Moreover, the presented implementation details highlight the main methods that might need modifications to implement more specific functionality or emulate other similar access methods (\textit{e.g.}, OFDMA).
\end{abstract}


\section{Introduction}
\label{sec:introduction}
\sloppy{
When the medium needs to be shared between multiple active users, the resource is fairly distributed with multiple access techniques, such as Multi-Frequency Time Division Multiple Access (MF-TDMA) or Orthogonal Frequency-Division Multiple Access (OFDMA). OFMDA is used in 4G/LTE and WiMAX/802.16 whilst MF-TDMA is used in Digital Video Broadcasting (DVB). Considering recent deployments of such multiple access networks, it is important to study the interactions of these access methods (PHY/MAC) with the rest of the network protocols (transport layer) in order to ensure optimal experience for the end user. This can be achieved through large simulations studies.
}


Several modules simulating OFDMA or MF-TDMA techniques for NS-2~\cite{ns2_ref} already exist. In the context of OFDMA, WiMAX has seen a lot of interest~\cite{ns2_wimax_1,ns2_wimax_2}. The current DVB-S2/RCS specifications have also been implemented with MF-TDMA characteristics~\cite{ns2_laas,ns2_aberdeen}. These modules attempt to be as close as possible to the real systems and the layout of their components, which make them unsuitable to assess proposed changes to the DVB-S2/RCS architecture (\textit{e.g.}, access methods strategies), nor to extend their work to other MF-TDMA networks. Morever, this level of realism might not be necessary for the study of high layer behaviours. Indeed, Gurtov and Floyd claim that a better trade-off between generality, realism and accurate modeling can be found to improve transport protocol performance evaluation~\cite{floyd_gurtov}.

We therefore propose an NS-2 module, Physical Channel Access (PCA), which emulates packet delays at the access point based on specific access methods parameters. This module allows to integrate channel access considerations within NS-2 and assess their impact on upper layers performance. We follow the idea of~\cite{floyd_gurtov} by emulating the characteristics of various physical channel access techniques rather than extending MAC/PHY simulation models. This therefore allows to conveniently study a wider range of scenarios than the modules presented in~\cite{ns2_laas,ns2_aberdeen}:
\begin{itemize}
\item experimental channel access strategies for MF-TDMA based systems (\textit{e.g.}, current drafts);
\item generic time-frequency multiplexing architectures;
\item adaptive access methods.
\end{itemize}
PCA is well suited for current needs in the extension of the DVB specifications: it is based on the currently standardized parameters but allows to depart from them for further investigations. For example, it allows to consider experimental access methods and capacity allocation processes which are under discussion to support transmission of home user data on the satellite return channel (RCS), so far reserved to signaling. 

The rest of the paper is organized as follows. In Section~\ref{sec:integration_in_ns2}, we present the general concepts behind MF-TDMA and notations relevant to the rest of this paper. We describe the implementation of PCA in Section~\ref{sec:interaction_between_component}. We document the internal parameters and present a use case in the context of DVB-RCS2 in Section~\ref{sec:simulation_tcl}. We conclude and discuss future work in Section~\ref{sec:conclusion}.   

\section{MF-TDMA networks}
\label{sec:integration_in_ns2}
On an MF-TDMA link, the capacity is shared at the \textit{Access Point}: it is dynamically distributed on times $\times$ frequency blocks (denoted ``frame'' in this article). The access point (the NS-2 node where the module we present in Section~\ref{sec:interaction_between_component} is introduced) forwards traffic from one or more users to one or more receivers over the shared medium, therefore covering both up and down link scenarios.

Before presenting in details our NS-2 module, we provide some definitions of the terms used in this paper:
\begin{itemize}
	\item Flow: data transfer at the transport layer;
	\item Datagram: network layer segment of a flow;
	\item Link Layer Data Unit (LLDU): $N_\mathrm{data}$ bytes of a fragmented datagram;
	\item Physical Layer Data Unit (PLDU): LLDU with an optional $N_\mathrm{repair}$ recovery bytes ($N=N_\mathrm{data}+N_\mathrm{repair}$);
	\item Block: PLDUs can be further split into $N_\mathrm{block}$ blocks if the access method requires; 
	\item Slots: element of a frame where a block can be scheduled.
\end{itemize}

\subsection{Access methods}
In the context of MF-TDMA networks, two classes of access methods can be introduced: dedicated and random access methods. 

\paragraph{Dedicated access} 
When a new flow arrives at the access point, parts of the channel have to be reserved for it. This induces a delay resulting from the reservation negociation process. The reservation ensures that capacity is fairly distributed: if there are 40 slots available and 10 users, each user can transmit data on 4 slots.

\paragraph{Random access}
No reservation is needed and data can be transmitted without additional delay. However several users can unknowingly use the same slot and data risks not to be recovered. Stronger error codes are introduced at the physical layer and each user can transmit a reduced $N_\mathrm{data}$ useful bytes: $N_\mathrm{repair}$ redundancy bytes are added to the $N_\mathrm{data}$ bytes to form a code word of $N=N_\mathrm{data}+N_\mathrm{repair}$ bytes that are split into $N_\mathrm{block}$ blocks. $N_\mathrm{ra}$ slots form a \textit{Random Access} block (RA block) on which erasure codes are introduced. Each transmitter randomly spreads its $N_\mathrm{block}$ blocks across the $N_\mathrm{ra}$ slots of the RA block for spectral diversity. 

Performance of random access methods can be described by the probability that a receiver decodes its $N_\mathrm{data}$ useful bytes depending on the number of users that transmit data on the RA block. Table~\ref{tab:perf_random_access} shows a generic example of such a description where $P_{i,j}$ is the probability that a packet cannot be recovered by the receiver when there are $N_U \in [NbUser_{j};NbUser_{j+1}]$ users on the RA block and and the signal-to-noise ratio of the channel is $Es/N0_{i}$.

\begin{table}[h]
\caption{Random access method performance}
\begin{center}
\scalebox{0.8}{
\begin{tabular}{ccccccc}
\toprule
0 & $NbUser_{1}$ & $NbUser_{2}$ & $NbUser_{3}$ & $\dots$ & $NbUser_{26}$ & $0$ \\
$Es/N0_{1}$ & $P_{1,1}$ & $P_{1,2}$ & $P_{1,3}$ & $\dots$ & $P_{1,26}$ & $0$ \\
$Es/N0_{2}$ & $P_{2,1}$ & $P_{2,2}$ & $P_{2,3}$ & $\dots$ & $P_{2,26}$ & $0$ \\
$\dots$ & $\dots$ & $\dots$ & $\dots$ & $\dots$ & $\dots$ & $\dots$ \\
$Es/N0_{X}$ & $P_{X,1}$ & $P_{X,2}$ & $P_{X,3}$ & $\dots$ & $P_{X,26}$ & $0$ \\ 
\bottomrule
\end{tabular}}
\end{center}
\label{tab:perf_random_access}
\end{table}

\subsection{Frame structure}
\begin{figure}[h!]
	\includegraphics[width=\linewidth]{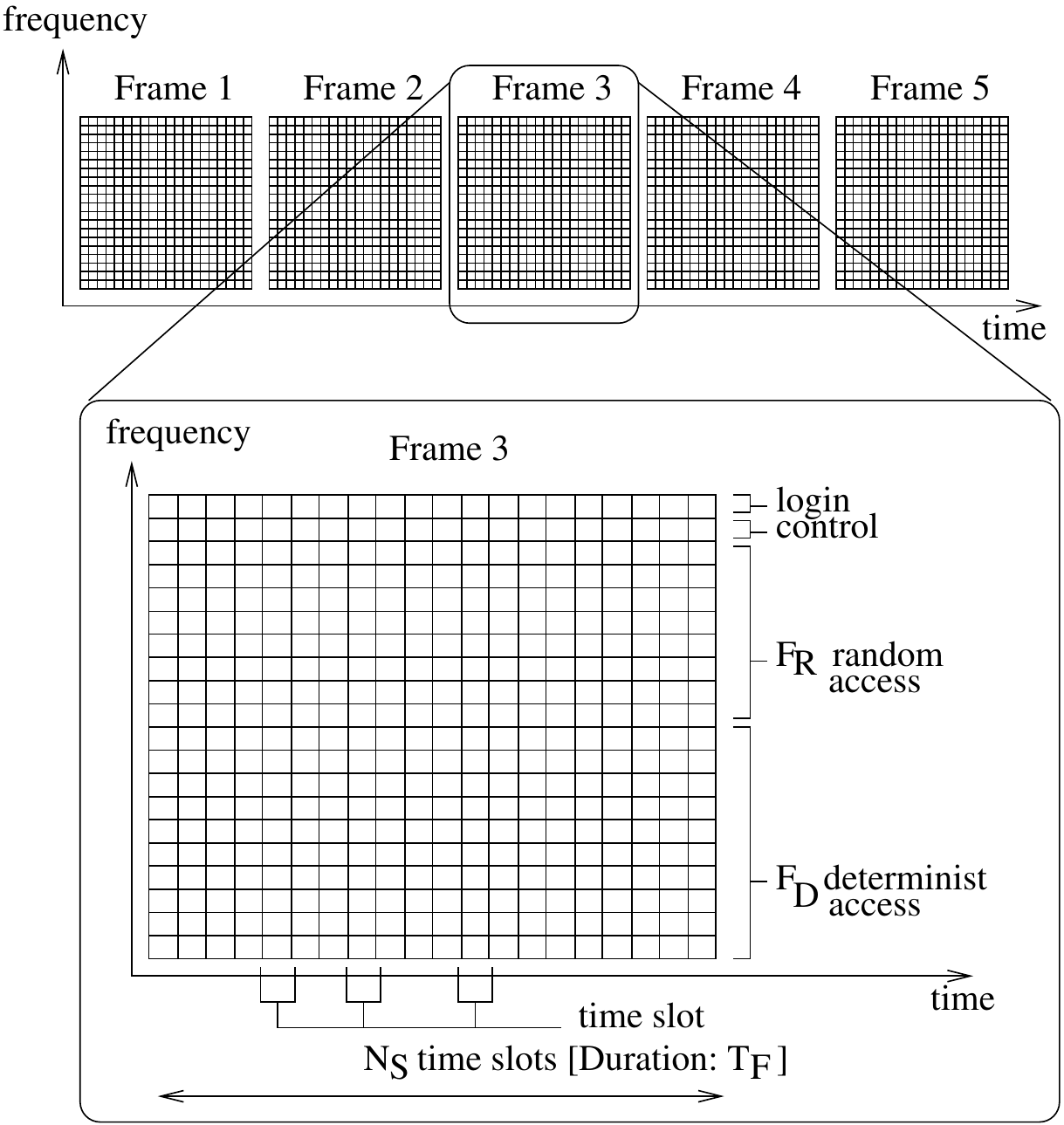}
	\caption{Times $\times$ Frequency block description}
	\label{fig:frame_structure}
\end{figure}

The capacity is dynamically distributed between the different users on the time $\times$ frequency frame which structure is detailed in Figure~\ref{fig:frame_structure}. At the access point, transmission of a frame is scheduled every $T_{F}$. We denote by $N_{S}$ the number of time slots available per frequency. The frequencies on which data is transmitted can be divided depending on the access method: $F_{R}$ frequencies are dedicated to the random access methods and $F_{D}$ are reserved to the dedicated access methods. In total, a frame can carry $N_{S} \times (F_{R}+F_{D})$ slots. 

\subsection{Antenna limitations}
It is possible that some transmitters cannot send data on different frequencies at once. This limitation has to be considered when determining the maximum number of slots that a user is allowed to occupy on each frame. This is highly linked to the frame structure, and in this case, a flow can only use $N_{S}$ slots whatever the number of available frequencies. 

As an example, if $N_{S}=40$\,slots and:
\begin{itemize}
	\item $F_{R}=0~\&~F_{D}>1$ (dedicated access): a unique user can exploit $N_{S}=40$ slots;
	\item $F_{R}=1~\&~F_{D}=0$ (random access), $N_\mathrm{block}=3$, $N_\mathrm{ra}=40$: a unique user can exploit $ \lfloor N_{S}/N_\mathrm{block} \rfloor =13$ slots.
\end{itemize}

\section{Implementation details}
\label{sec:interaction_between_component}
In order to control delays, PCA is implemented as a queueing delay. We inherit from the \verb_DropTail_ queue management scheme, of which our PCA sub-class redefines the methods used to process the packets. Each node uses the \texttt{enque()} and \texttt{deque()} methods to add and remove packets from the queue. 

\begin{figure}[h!]
	\resizebox{\columnwidth}{!}{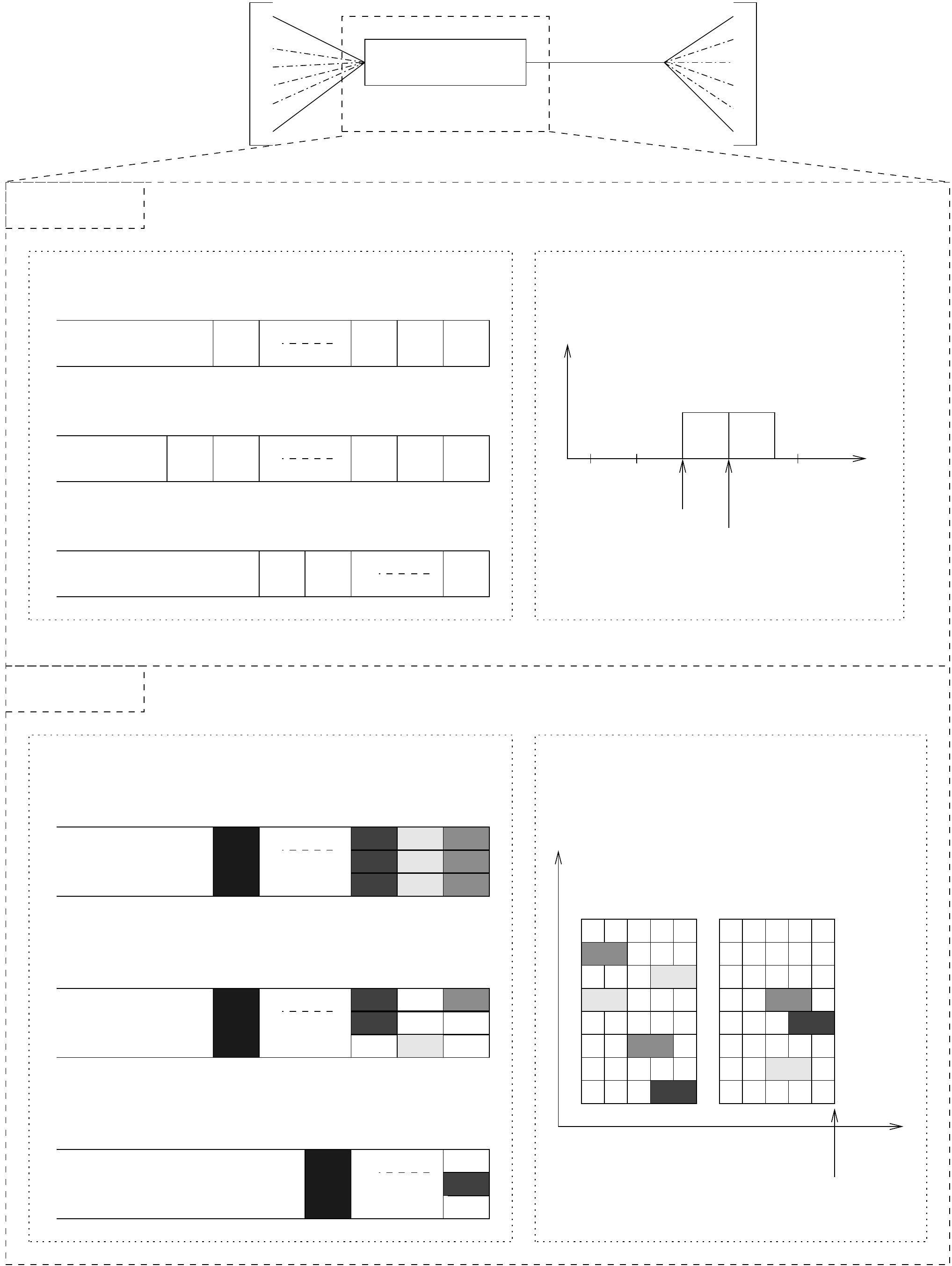}
	\caption{Capacity allocation: \texttt{enque()} and \texttt{deque()}}
	\label{fig:global_scheme_pca}
\end{figure}

In Figure~\ref{fig:global_scheme_pca}, we compare the \texttt{enque()} and \texttt{deque()} methods of \verb_DropTail_ and \verb_DropTail/PCA_. With \verb_DropTail_, when the \texttt{enque()} method adds packet $P_{N+1}$, it is added at the end of the sending buffer and transmitted when $P_{1}, \dots, P_{N}$ have been transmitted with the \texttt{deque()} method. With \verb_DropTail/PCA_, when a packet is \verb_enque()_ed, it is also added to the sending buffer. However, depending on the access method introduced, only a subset of the datagram is considered sent with each frame. When the last byte of a datagram has been transmitted, \texttt{deque()}, which is called every $T_{F}$, removes the packets from the sending buffer and passes it along.    

The \verb_DropTail/PCA_ queuing policy is implemented in two files (\texttt{pca.cc} and \texttt{pca.h}) located in the \texttt{queue/} sub-directory of the NS-2 source. We detail the content of these files  here. 

\subsection{Data structures}
Our module implements linked-lists in order to store information about the current flows and their packets. 

\paragraph{Packets list}
The packet list contains information about the different packets that have reached the access point node but have not been fully transmitted yet. Each packet is defined by: 
\begin{itemize}
	\item \verb#appl_id#: identifier of the flow;
	\item \verb#pkt_seqno#: sequence number in the flow;
	\item \verb#frame_in#: emulated frame number after which the data of the packet start to be transmitted;
	\item \verb#frame_out#: frame number at which the last bit of the packet must be transmitted;
	\item \verb#bool_first_frame#: boolean to specify if the connection needs to be established (first packet of the current application);
	\item \verb#bool_lost#: boolean specifying whether the packet is lost;
	\item \verb#bool_rand#: boolean specified if the access method is random (\verb#bool_rand#=1) or dedicated (\verb#bool_rand#=0);
	\item \verb#bits_to_send#: actual number of bits of the datagram that have not been sent yet;
	\item \verb#bits_next_frame#: number of bits that will be sent at the next frame;
	\item \verb#remaining_slot_frame_appl_det#: number of dedicated slots that remain for this packet's flow;
	\item \verb#remaining_slot_frame_appl_rnd#: number of random slots that remains for the flow of the packet;
	\item \verb#used_slot_frame_appl_rnd#: number of slots that the packet's flow will use in the next frame.
\end{itemize}

\paragraph{Applications list}
This linked list is used to collect information relative to the currently active applications. It tracks all the open connections and maintains information about the last transmitted datagrams.
\begin{itemize}
	\item \verb#appl_id#: identifier of the application;
	\item \verb#pkt_seq#: sequence number of the last packet transmitted;
	\item \verb#last_time_out#: time when the last packet of the given application has been sent.
\end{itemize}

\subsection{\texttt{enque(Packet *p)} method}
\begin{figure}[h!]
	\includegraphics[width=\linewidth]{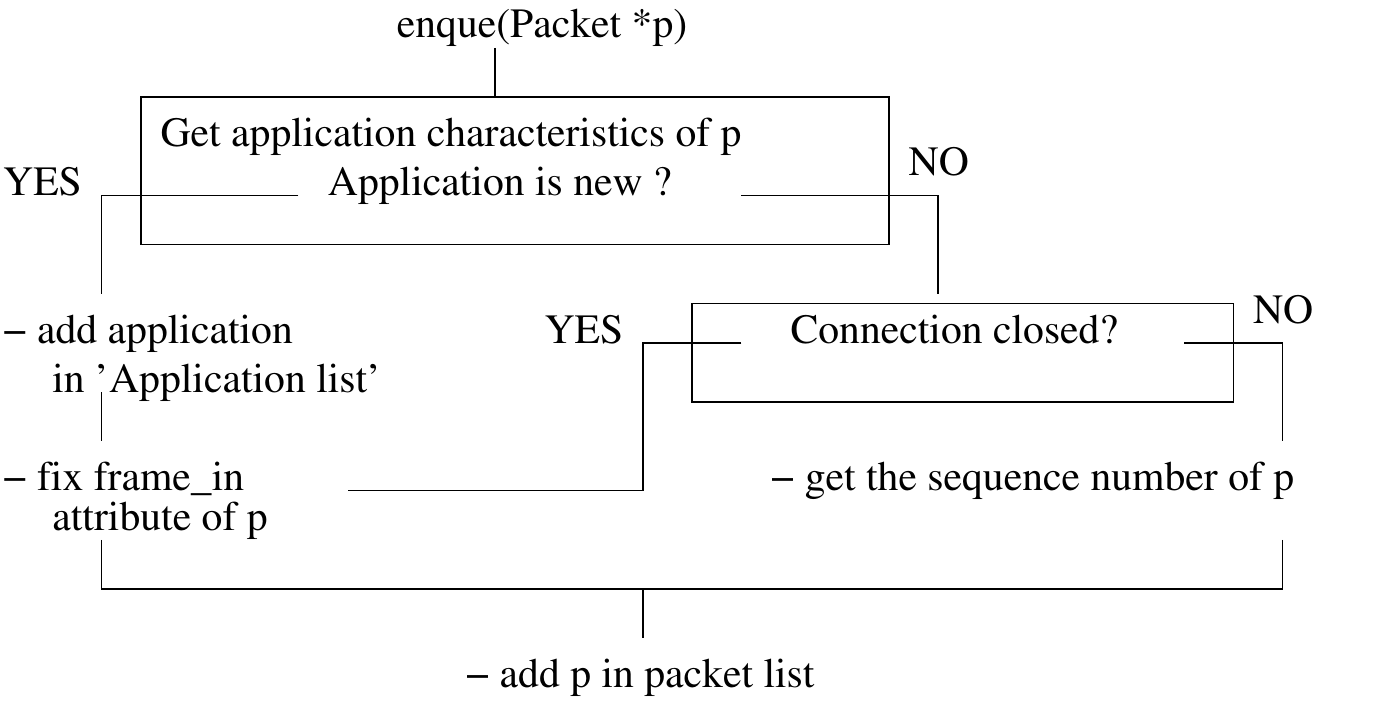}
	\caption{\texttt{enque()} method flowchart}
	\label{fig:enque_diagram}
\end{figure}

The \verb#enque()# method is called when the network layer passes a packet down to PCA. It registers the packet and its attributes for consideration in the capacity distribution process. Figure~\ref{fig:enque_diagram} summarises the operation of this function.
 
Before enqueuing a packet, it verifies whether the connection needs to be established and the application added in the applications list. If the application is not new, it checks whether the connection is still open. It then adds the packet with its characteristics to the packets list. If \verb#pkt_seqno#=$0$, it sets the \verb#frame_in# attribute of the packet (first frame where data from this packet starts being transmitted) depending on the access method. \verb#frame_out# is initially set to $\infty$, and later updated by the \verb#adaptBitNextFrame()# method described in the next section.
 
\subsection{\texttt{deque()} method}
\begin{figure}[h!]
	\includegraphics[width=\linewidth]{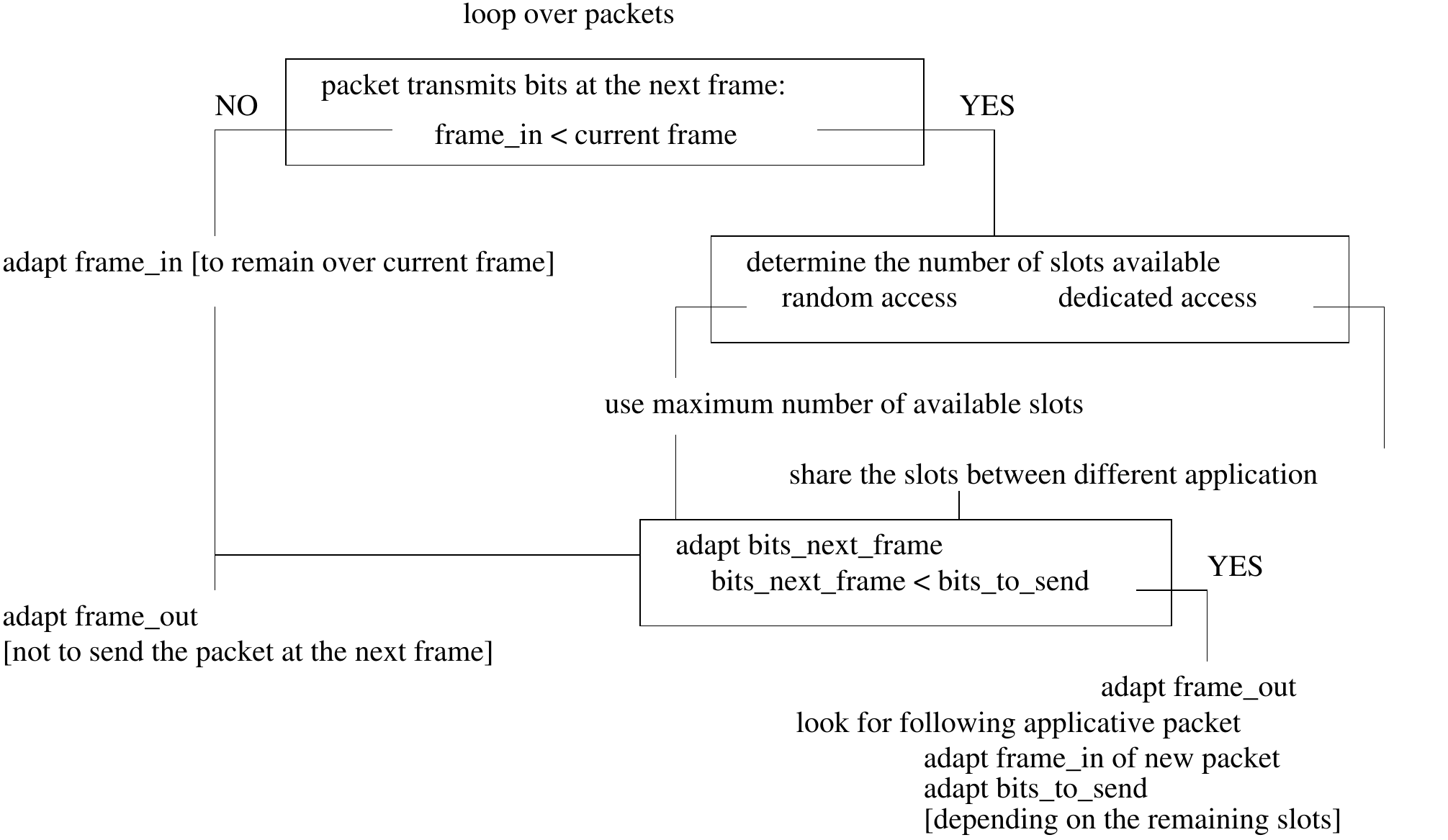}
	\caption{\texttt{adaptBitNextFrame()} method flowchart}
	\label{fig:adapt_bits_next_frame_diagram}
\end{figure}

The \verb^deque()^ method emulates arrival of a new frame at the receiver. It is called by a timer every $T_{F}$. It loops over the packet list, forwards the packet for which \verb#frame_out# is the current frame to the receiving node and updates the transmission progress of the other datagrams by adjusting their attributes. Figure~\ref{fig:adapt_bits_next_frame_diagram} details how the number of bits to transmit on the next frame is calculated in the \verb#adaptBitNextFrame()# function. This process has been introduced to take care of: 
\begin{itemize}
	\item fair distribution of the capacity with dedicated access methods;
	\item determination of erasure probability with random access methods (depending on the load of the link and methods performance as detailed in Table~\ref{tab:perf_random_access});
	\item adaptation of the packet transmission to ensure that flows send their packets in the order they have been received.
\end{itemize}

To do so, \verb#adaptBitNextFrame()# updates the values of \verb#bits_next_frame#, \verb#used_slot_frame_appl_rnd# and \verb#frame_out# as follows. At frame $F$, for each packet where \verb#frame_in#$ < F$, we compute $B_\mathrm{remaining}$=\verb#bits_to_send#-\verb#bits_next_frame#, which corresponds to the data which remains to be transmitted. If $B_\mathrm{remaining}>0$, the packet is left in the queue; \verb#bits_next_frame#, which corresponds to the data that will be transmitted at frame $F+1$, is determined depending on the access method (as well as the number of slots which it will use, \verb#used_slot_frame_appl_rnd#) and \verb#bits_to_send# is set to $B_\mathrm{remaining}$. If $B_\mathrm{remaining} \le 0$, \verb#frame_out# is set to $F+1$; the next packet for that application is then found in the packet list, its \verb#frame_in# is set to $F$ and $B_\mathrm{remaining}$ is subtracted from its \verb#bits_to_send#.

\subsection{Limitations and extensions}
PCA can be used to conduct large studies on (MF-)TDMA schemes, however it has some limitations. First, the performance of random access methods depends on the signal-to-noise ratio of the specific link between one receiver and the access point. It is currently assumed that this value is the same for all receivers, but this can be easily lifted by adapting the receiver-to-SNR mapping code. Second, PCA does not consider prioritization between flows. Nonetheless, this could be achieved by flagging packets at higher layers and inspecting these flags in \texttt{deque()} and \texttt{adaptBitNextFrame()} functions. 

The developement of this module has been driven by (MF-)TDMA specifications. However, it can easily be extended for other similar access methods (time and/or frequency multiplexing) by adapting the \texttt{adaptBitNextFrame()} method to reflect the specific data scheduling scheme of the desired technique. Also, in the current implementation, it was considered that one flow could only send a limited amount of data per RA block. This quantity can be adjusted in the simulation parameters (through \verb#sizeSlotRandom_# and \verb#nbSlotRndFreqGroup_#).  

\section{Use case example}
\label{sec:simulation_tcl}
In this section, we detail the principal parameters of the \verb#DropTail/PCA# queuing policy and illustrate them with an example in the context of DVB-RCS2. 

\subsection{Parameters}
The parameters are set following the standard NS-2 fashion: 

\verb#Queue/DropTail/PCA set <PARAMETER> <VALUE>#

The following parameters have to be specified prior to starting a simulation: 
\begin{itemize}
	\item \verb#cutConnect_#: time after which the connection between the gateway and the user is closed (in seconds);
	\item \verb#esN0_#: signal-to-noise ratio of the channel in dB (for random access methods performance);
	\item \verb#switchAleaDet_#: sequence number at which the access method switches from random to dedicated;
	\item \verb#frameDuration_# ($T_{F}$): duration of a frame;
	\item \verb#nbSlotPerFreq_# ($N_{S}$): number of time slots per frequency;
	\item \verb#sizeSlotRandom_# ($N_\mathrm{data}$): useful number of bits that can be sent on one RA block (\textit{i.e.}, where random access methods are introduced);
	\item \verb#sizeSlotDeter_# ($N_\mathrm{data}$): useful number of bits for each time slots where dedicated access methods are introduced;
	\item \verb#rtt_#: two-way link delay (in seconds);
	\item \verb#freqRandom_# ($F_{R}$): number of frequencies used for random access;
	\item \verb#nbFreqPerRand_# ($(F_{R} \times N_{S}) / N_\mathrm{ra}$): number of frequencies comprised in an RA block;
	\item \verb#freqDeter_# ($F_{D}$): number of frequencies used for dedicated access;
	\item \verb#maxThroughtput_#: maximum authorized throughput for one given flow (in Mbps);
	\item \verb#nbSlotRndFreqGroup_# ($N_\mathrm{block}$): number of blocks a PLDU is split into for distribution in one RA block;
	\item \verb#boolAntennaLimit_#: boolean whether one transmitter has one or $F_{R}+F_{D}$ antennas.
\end{itemize}

In order to introduce PCA, the link between two nodes N1 and N2 (N1 is the access point node) can then be defined as: 
\begin{verbatim}
$ns simplex-link \ 
	$N1 $N2 $bandwidth [$rtt_ / 2]  \ 
	DropTail/PCA $random_access_file_performance
\end{verbatim}
where \verb#$random_access_file_performance# is the name of the file containing information about random access performance laid out as in Table~\ref{tab:perf_random_access}.  

\subsection{Use case in the context of DVB-RCS2}
We illustrate the results obtained with this PCA module in the context of DVB-RCS2. DVB-RCS2 is the return link on which the home user transmits data to the satellite gateway: the satellite link is shared between the different users. There is a recent interest in enabling the home user to transmit data (\textit{e.g.} for web browsing or email exchange) through this channel. However, there is contention about which access method is best suited for this use. PCA allows to provide insight on this question by evaluting transport layer performance with various access method proposals. 

We consider three different cases: (1) a dedicated access method and random access methods ((2) CRDSA~\cite{ref_crdsa} and (3) MuSCA~\cite{ref_musca}). We use input for each random access method that follows the format illustrated in Section~\ref{sec:integration_in_ns2}. We base the choice of parameters on specifications defined in~\cite{dvb_RCS2_norm} and present them in Table~\ref{tab:dvb_param}. 

\begin{table}[h]
\caption{Use case simulation parameters}
\begin{center}
\scalebox{0.8}{
\begin{tabular}{c|ccc}
\toprule
Parameters & & Access method & \\
 & Dedicated & Random & Random \\
 & & (CRDSA) & (MUSCA) \\
\cmidrule{1-4}
\texttt{cutConnect\_} & 3 & 3 & 3 \\
\texttt{esN0\_} & 5 & 5 & 5 \\
\texttt{switchAleaDet\_} & 0 & $\infty$ & $\infty$ \\
\texttt{frameDuration\_} & 0.045 & 0.045 & 0.045 \\
\texttt{nbSlotPerFreq\_} & 40 & 40 & 40 \\
\texttt{sizeSlotRandom\_} & xx & 613 & 680 \\
\texttt{sizeSlotDeter\_} & 920 & xx & xx \\
\texttt{rtt\_} & 0.5 & 0.5 & 0.5 \\
\texttt{freqRandom\_} & 0 & 100 & 100 \\
\texttt{nbFreqPerRand\_} & 2.5 & 2.5 & 2.5 \\
\texttt{freqDeter\_} & 100 & 0 & 0 \\
\texttt{maxThroughtput\_} & 1\,Mbps & 1\,Mbps & 1\,Mbps \\
\texttt{nbSlotRndFreqGroup\_} & xx & 3  & 3 \\
\texttt{boolAntennaLimit\_} & 1 & 1 & 1 \\
\bottomrule
\end{tabular}}
\end{center}
\label{tab:dvb_param}
\end{table}

We consider two nodes in NS-2. The first node transmits a various number of FTP flows to the second node. This allows to study the transport layer performance (no data-starved sender). The size of IP packets is 1500\,bytes, and the queue at the sender is large enough not to be overflowed. We use the Linux implementation of TCP Reno, with SACK options. The simulation time is 20\,seconds. 

\begin{figure}[h!]
	\includegraphics[width=\linewidth]{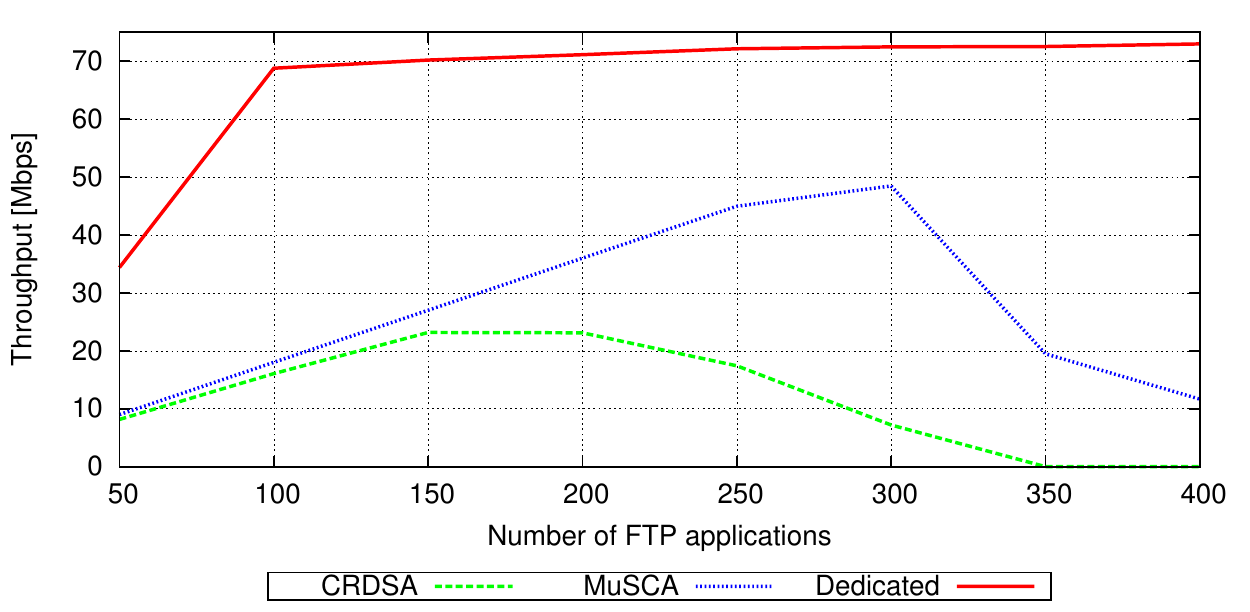}
	\caption{Throughput depending on the load of the network}
	\label{fig:through_use_case}
\end{figure}

In Figure~\ref{fig:through_use_case}, we show the total achievable throughput measured on the shared link level. This shows that dedicated access methods support more load on the network whereas random access methods, which PCA allows us to study, experience lower throughput with increasing loads. 

\begin{figure}[h!]
	\includegraphics[width=\linewidth]{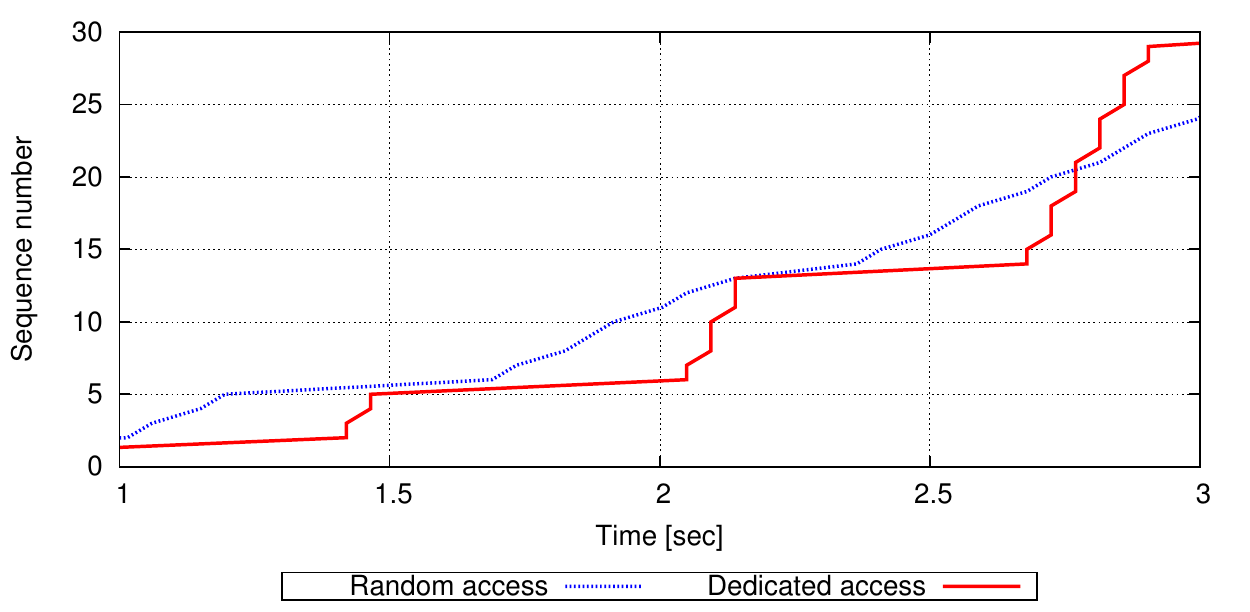}
	\caption{Sequence number evolution}
	\label{fig:cwnd_use_case}
\end{figure}

Figure~\ref{fig:cwnd_use_case} illustrates the evolution of the packet sequence numbers of one given FTP application during the first seconds of the simulation. We only show the results with MuSCA as the random access method; the performance with CRDSA is qualitatively similar. Contrary to the previous metrics, random access methods seem to perform better. Indeed, thanks to a faster connection establishment, random access methods transmit the first packets faster than dedicated access methods. However, with dedicated access, the time needed to effectively transmit one packet is smaller: as detailed in Table~\ref{tab:dvb_param}, more bits can be sent on one slot (\textit{i.e.}, \verb#sizeSlotDeter_#$>$\verb#sizeSlotRandom_#).    

These preliminary simulations present a first evaluations of the impact of access methods on TCP performance in the context of DVB-RCS2. We expect to explore and analyze this use case further.

\section{Conclusion and future work}
\label{sec:conclusion}
In this article, we presented PCA, a module for NS-2 that enables to emulate channel access methods for the evaluation of the interaction with transport protocol mechanisms. This module considers time and/or frequency multiplexing access methods and can be used in contexts where the capacity of the channel is shared among multiple users. We detailed the main components and their operation as well as the parameters required to configure the module.

PCA is useful to assess the impact of medium access strategies on transport performance, especially for satellite links. It was initially developed with MF-TDMA specifications in mind, however we also indicated where extensions should be made to accomodate other similar technologies (\textit{e.g.}, OFDMA). 

We are currently using this module to investigate random access performance in the context of DVB-RCS2. Also, we plan to release the module as open source soon.

\end{document}

%% file: global_scheme_pca.pdftex_t
\begin{picture}(0,0)%
\includegraphics{global_scheme_pca.pdf}%
\end{picture}%
\setlength{\unitlength}{4144sp}%
\begingroup\makeatletter\ifx\SetFigFont\undefined%
\gdef\SetFigFont#1#2#3#4#5{%
  \reset@font\fontsize{#1}{#2pt}%
  \fontfamily{#3}\fontseries{#4}\fontshape{#5}%
  \selectfont}%
\fi\endgroup%
\begin{picture}(9294,12354)(3319,-13573)
\put(10171,-5506){\makebox(0,0)[b]{\smash{{\SetFigFont{12}{14.4}{\rmdefault}{\mddefault}{\updefault}{\color[rgb]{0,0,0}$P_1$}%
}}}}
\put(10621,-5506){\makebox(0,0)[b]{\smash{{\SetFigFont{12}{14.4}{\rmdefault}{\mddefault}{\updefault}{\color[rgb]{0,0,0}$P_2$}%
}}}}
\put(8821,-12436){\makebox(0,0)[b]{\smash{{\SetFigFont{12}{14.4}{\rmdefault}{\mddefault}{\updefault}{\color[rgb]{0,0,0}$T_1$}%
}}}}
\put(10171,-12436){\makebox(0,0)[b]{\smash{{\SetFigFont{12}{14.4}{\rmdefault}{\mddefault}{\updefault}{\color[rgb]{0,0,0}$T_2$}%
}}}}
\put(11521,-12436){\makebox(0,0)[b]{\smash{{\SetFigFont{12}{14.4}{\rmdefault}{\mddefault}{\updefault}{\color[rgb]{0,0,0}$T_3$}%
}}}}
\put(11116,-5911){\makebox(0,0)[b]{\smash{{\SetFigFont{12}{14.4}{\rmdefault}{\mddefault}{\updefault}{\color[rgb]{0,0,0}$T_3$}%
}}}}
\put(9091,-5911){\makebox(0,0)[b]{\smash{{\SetFigFont{12}{14.4}{\rmdefault}{\mddefault}{\updefault}{\color[rgb]{0,0,0}$T_1$}%
}}}}
\put(7426,-5731){\makebox(0,0)[b]{\smash{{\SetFigFont{12}{14.4}{\rmdefault}{\mddefault}{\updefault}{\color[rgb]{0,0,0}$P_2$}%
}}}}
\put(5626,-4606){\makebox(0,0)[b]{\smash{{\SetFigFont{12}{14.4}{\rmdefault}{\mddefault}{\updefault}{\color[rgb]{0,0,0}$P_N$}%
}}}}
\put(6976,-4606){\makebox(0,0)[b]{\smash{{\SetFigFont{12}{14.4}{\rmdefault}{\mddefault}{\updefault}{\color[rgb]{0,0,0}$P_3$}%
}}}}
\put(7426,-4606){\makebox(0,0)[b]{\smash{{\SetFigFont{12}{14.4}{\rmdefault}{\mddefault}{\updefault}{\color[rgb]{0,0,0}$P_2$}%
}}}}
\put(7876,-4606){\makebox(0,0)[b]{\smash{{\SetFigFont{12}{14.4}{\rmdefault}{\mddefault}{\updefault}{\color[rgb]{0,0,0}$P_1$}%
}}}}
\put(5176,-5731){\makebox(0,0)[b]{\smash{{\SetFigFont{12}{14.4}{\rmdefault}{\mddefault}{\updefault}{\color[rgb]{0,0,0}$P_{N+1}$}%
}}}}
\put(5626,-5731){\makebox(0,0)[b]{\smash{{\SetFigFont{12}{14.4}{\rmdefault}{\mddefault}{\updefault}{\color[rgb]{0,0,0}$P_N$}%
}}}}
\put(6976,-5731){\makebox(0,0)[b]{\smash{{\SetFigFont{12}{14.4}{\rmdefault}{\mddefault}{\updefault}{\color[rgb]{0,0,0}$P_3$}%
}}}}
\put(7876,-5731){\makebox(0,0)[b]{\smash{{\SetFigFont{12}{14.4}{\rmdefault}{\mddefault}{\updefault}{\color[rgb]{0,0,0}$P_1$}%
}}}}
\put(6076,-6856){\makebox(0,0)[b]{\smash{{\SetFigFont{12}{14.4}{\rmdefault}{\mddefault}{\updefault}{\color[rgb]{0,0,0}$P_{N+1}$}%
}}}}
\put(6526,-6901){\makebox(0,0)[b]{\smash{{\SetFigFont{12}{14.4}{\rmdefault}{\mddefault}{\updefault}{\color[rgb]{0,0,0}$P_N$}%
}}}}
\put(7876,-6856){\makebox(0,0)[b]{\smash{{\SetFigFont{12}{14.4}{\rmdefault}{\mddefault}{\updefault}{\color[rgb]{0,0,0}$P_3$}%
}}}}
\put(5626,-9106){\makebox(0,0)[b]{\smash{{\SetFigFont{12}{14.4}{\rmdefault}{\mddefault}{\updefault}{\color[rgb]{0,0,0}$P_N$}%
}}}}
\put(6976,-9106){\makebox(0,0)[b]{\smash{{\SetFigFont{12}{14.4}{\rmdefault}{\mddefault}{\updefault}{\color[rgb]{0,0,0}$P_3$}%
}}}}
\put(7426,-9106){\makebox(0,0)[b]{\smash{{\SetFigFont{12}{14.4}{\rmdefault}{\mddefault}{\updefault}{\color[rgb]{0,0,0}$P_2$}%
}}}}
\put(7876,-9106){\makebox(0,0)[b]{\smash{{\SetFigFont{12}{14.4}{\rmdefault}{\mddefault}{\updefault}{\color[rgb]{0,0,0}$P_1$}%
}}}}
\put(5626,-10681){\makebox(0,0)[b]{\smash{{\SetFigFont{12}{14.4}{\rmdefault}{\mddefault}{\updefault}{\color[rgb]{0,0,0}$P_N$}%
}}}}
\put(6976,-10681){\makebox(0,0)[b]{\smash{{\SetFigFont{12}{14.4}{\rmdefault}{\mddefault}{\updefault}{\color[rgb]{0,0,0}$P_3$}%
}}}}
\put(7426,-10681){\makebox(0,0)[b]{\smash{{\SetFigFont{12}{14.4}{\rmdefault}{\mddefault}{\updefault}{\color[rgb]{0,0,0}$P_2$}%
}}}}
\put(7876,-10681){\makebox(0,0)[b]{\smash{{\SetFigFont{12}{14.4}{\rmdefault}{\mddefault}{\updefault}{\color[rgb]{0,0,0}$P_1$}%
}}}}
\put(6526,-12256){\makebox(0,0)[b]{\smash{{\SetFigFont{12}{14.4}{\rmdefault}{\mddefault}{\updefault}{\color[rgb]{0,0,0}$P_N$}%
}}}}
\put(7876,-12256){\makebox(0,0)[b]{\smash{{\SetFigFont{12}{14.4}{\rmdefault}{\mddefault}{\updefault}{\color[rgb]{0,0,0}$P_3$}%
}}}}
\put(3826,-12211){\makebox(0,0)[lb]{\smash{{\SetFigFont{12}{14.4}{\rmdefault}{\mddefault}{\updefault}{\color[rgb]{0,0,0}at $T_3$}%
}}}}
\put(3826,-10636){\makebox(0,0)[lb]{\smash{{\SetFigFont{12}{14.4}{\rmdefault}{\mddefault}{\updefault}{\color[rgb]{0,0,0}at $T_2$}%
}}}}
\put(3826,-9061){\makebox(0,0)[lb]{\smash{{\SetFigFont{12}{14.4}{\rmdefault}{\mddefault}{\updefault}{\color[rgb]{0,0,0}at $T_1$}%
}}}}
\put(3601,-7981){\makebox(0,0)[lb]{\smash{{\SetFigFont{12}{14.4}{\rmdefault}{\mddefault}{\updefault}{\color[rgb]{0,0,0}PCA}%
}}}}
\put(3826,-6361){\makebox(0,0)[lb]{\smash{{\SetFigFont{12}{14.4}{\rmdefault}{\mddefault}{\updefault}{\color[rgb]{0,0,0}at $T_3$}%
}}}}
\put(3826,-5236){\makebox(0,0)[lb]{\smash{{\SetFigFont{12}{14.4}{\rmdefault}{\mddefault}{\updefault}{\color[rgb]{0,0,0}at $T_2$ : enque($P_{N+1}$)}%
}}}}
\put(3601,-3211){\makebox(0,0)[lb]{\smash{{\SetFigFont{12}{14.4}{\rmdefault}{\mddefault}{\updefault}{\color[rgb]{0,0,0}DROPTAIL}%
}}}}
\put(4951,-3886){\makebox(0,0)[lb]{\smash{{\SetFigFont{12}{14.4}{\rmdefault}{\mddefault}{\updefault}{\color[rgb]{0,0,0}ACCESS POINT BUFFER}%
}}}}
\put(3826,-4111){\makebox(0,0)[lb]{\smash{{\SetFigFont{12}{14.4}{\rmdefault}{\mddefault}{\updefault}{\color[rgb]{0,0,0}at $T_1$}%
}}}}
\put(4951,-8611){\makebox(0,0)[lb]{\smash{{\SetFigFont{12}{14.4}{\rmdefault}{\mddefault}{\updefault}{\color[rgb]{0,0,0}ACCESS POINT BUFFER}%
}}}}
\put(9676,-8611){\makebox(0,0)[lb]{\smash{{\SetFigFont{12}{14.4}{\rmdefault}{\mddefault}{\updefault}{\color[rgb]{0,0,0}SHARED LINK}%
}}}}
\put(11296,-13111){\makebox(0,0)[lb]{\smash{{\SetFigFont{12}{14.4}{\rmdefault}{\mddefault}{\updefault}{\color[rgb]{0,0,0}deque($P_2$)}%
}}}}
\put(11296,-12886){\makebox(0,0)[lb]{\smash{{\SetFigFont{12}{14.4}{\rmdefault}{\mddefault}{\updefault}{\color[rgb]{0,0,0}deque($P_1$)}%
}}}}
\put(9991,-6586){\makebox(0,0)[lb]{\smash{{\SetFigFont{12}{14.4}{\rmdefault}{\mddefault}{\updefault}{\color[rgb]{0,0,0}deque($P_2$)}%
}}}}
\put(9541,-6361){\makebox(0,0)[lb]{\smash{{\SetFigFont{12}{14.4}{\rmdefault}{\mddefault}{\updefault}{\color[rgb]{0,0,0}deque($P_1$)}%
}}}}
\put(9451,-3886){\makebox(0,0)[lb]{\smash{{\SetFigFont{12}{14.4}{\rmdefault}{\mddefault}{\updefault}{\color[rgb]{0,0,0}SHARED LINK}%
}}}}
\put(10891,-2041){\makebox(0,0)[lb]{\smash{{\SetFigFont{12}{14.4}{\rmdefault}{\mddefault}{\updefault}{\color[rgb]{0,0,0}RECEIVERS}%
}}}}
\put(8776,-2041){\makebox(0,0)[lb]{\smash{{\SetFigFont{12}{14.4}{\rmdefault}{\mddefault}{\updefault}{\color[rgb]{0,0,0}SHARED}%
}}}}
\put(6931,-1861){\makebox(0,0)[lb]{\smash{{\SetFigFont{12}{14.4}{\rmdefault}{\mddefault}{\updefault}{\color[rgb]{0,0,0}ACCESS POINT}%
}}}}
\put(4276,-2041){\makebox(0,0)[lb]{\smash{{\SetFigFont{12}{14.4}{\rmdefault}{\mddefault}{\updefault}{\color[rgb]{0,0,0}SENDERS}%
}}}}
\put(8911,-2266){\makebox(0,0)[lb]{\smash{{\SetFigFont{12}{14.4}{\rmdefault}{\mddefault}{\updefault}{\color[rgb]{0,0,0}LINK}%
}}}}
\put(9541,-5911){\makebox(0,0)[b]{\smash{{\SetFigFont{12}{14.4}{\rmdefault}{\mddefault}{\updefault}{\color[rgb]{0,0,0}$T_2$}%
}}}}
\put(12151,-12436){\makebox(0,0)[b]{\smash{{\SetFigFont{12}{14.4}{\rmdefault}{\mddefault}{\updefault}{\color[rgb]{0,0,0}$t$}%
}}}}
\put(8866,-4561){\makebox(0,0)[b]{\smash{{\SetFigFont{12}{14.4}{\rmdefault}{\mddefault}{\updefault}{\color[rgb]{0,0,0}$f$}%
}}}}
\put(8776,-9511){\makebox(0,0)[b]{\smash{{\SetFigFont{12}{14.4}{\rmdefault}{\mddefault}{\updefault}{\color[rgb]{0,0,0}$f$}%
}}}}
\end{picture}%